\documentclass{elsart}

\usepackage{amsmath}
\usepackage{amssymb}
\usepackage[a4paper]{geometry}
\usepackage{graphicx}
\usepackage{epsfig}

\newcommand{\bsigma }{{\mbox{\boldmath $\sigma$}}}

\newcommand{\bphi   }{\mbox{\boldmath $\phi$}}

\newcommand{\bOmega }{\mbox{\boldmath $\Omega$}}

\newcommand{\bra    }{\langle}
\newcommand{\ket    }{\rangle}
\newcommand{\Bra    }{\left\langle}
\newcommand{\Ket    }{\right\rangle}

\newcommand{\rmd    }{\textrm{d}}

\begin{document}

\begin{frontmatter}

\title{Statics and dynamics of the Lebwohl-Lasher model in the Bethe
approximation}

\author{N.S.\@  Skantzos \thanksref{nikos}}
\address{Instituut voor Theoretische Fysica, Celestijnenlaan
200D, Katholieke Universiteit Leuven, B-3001, Belgium}
\and \author{J.P.L.\@ Hatchett \thanksref{jon}}
\address{$\dag$~ Hymans Robertson LLP, One London Wall, London EC2Y 5EA, UK}
\thanks[nikos]{nikos@itf.fys.kuleuven.be}
\thanks[jon]{jon.hatchett@hymans.co.uk}
\begin{abstract}
We study the Lebwohl-Lasher model for systems in which spin are
arranged on random graph lattices. At equilibrium our analysis
follows the theory of spin-systems on random graphs which allows us
to derive exact bifurcation conditions for the phase diagram. We
also study the dynamics of this model using a variant of the
dynamical replica theory. Our results are tested against
simulations. \vspace*{0.5cm}

\noindent
PACS : {89.75.-k, 75.10.Nr, 64.70.Md}
\end{abstract}
\end{frontmatter}

\maketitle

\section{Introduction}

Some physical systems are known to exhibit remarkably sharp, and yet
continuous, transitions between the ordered and paramagnetic phase.
The most prominent example of these are liquid crystals, systems
which combine order like that found in solids with fluidity like
that of liquids. Liquid crystals, generally modeled as systems of
`hard-rods', typically exhibit a transition between a phase with no
orientational or  translational order and a phase where an ordered
structure appears. The nature of this transition (continuous vs
discontinuous) is an important aspect in practical applications.

One of the most successful models that is able to capture the main
characteristics of this transition was introduced by Lebwohl and
Lasher in 1973 \cite{lebwohl}. In this model the microscopic degrees
of freedom $\phi_i$ take real values in the interval $[0,2\pi)$,
representing the orientation of rod $i$ relative to some fixed
reference point. To model the sharp transition, the coupling energy
between any pair of rods is taken to have the shape of a deep and
narrow well. This energy is taken to be $
\epsilon_{ij}(\phi_i-\phi_j)=-JL_p\left(\cos(\phi_i-\phi_j)\right) $
for any pair of rods $(i,j)$ where $J$ represents the strength of
the interaction and the function $L_p(x)$ denotes a $p$-th order
Legendre polynomial. In the context of `hard-rods', $p$ is taken to
be even to enforce invariance of the energy $\epsilon_{ij}$ under
the transformation $\phi\to\phi+\pi$. The value of $p$ plays a
crucial role in the nature of the phase transition
\cite{domany,jonsson}. Generally, this model has been the subject of
a significant amount of research (see for instance
\cite{domany,jonsson,priezjev2,berche,farinas2,selinger,zhang,fabbri}
and references therein).

In this paper we study the Lebwohl-Lasher model on a sparse graph
motivated both by the interesting nature of the transition and the
relative scarcity of analytic results. Our approach can be viewed as
the Bethe approximation to the finite dimensional problem, whereby
for every site there is an explicit local neighborhood. At
equilibrium our analysis follows the finite connectivity (as opposed
to fully-connected mean field) theory \cite{mezard} as developed for
real-valued spin systems \cite{XYsmallworld,CoolenXY}. In the
thermodynamic limit we are able to solve this model exactly and
derive expressions describing the bifurcation lines in the phase
diagram. Numerical evaluation of the order parameters agrees well
with the results of the bifurcation analysis and shows that the
phase transition, although sharp, is second-order. Comparison of the
above analytic results with simulation (Langevin) expreriments shows
excellent agreement.

We have also studied the dynamics of this model. Here, in contrast
to the thermodynamic analysis which follows a well-studied
framework, the terrain of the finite-connectivity dynamics is much
less explored. Some recent advances are the papers
\cite{GuzaiDRT,SemerjianWeigt,WeigtGoosens}. In the present study we
have chosen to extend the results of \cite{GuzaiDRT} in order to
account for the continuous nature of the spin variables. As in the
dynamical replica theory of \cite{Coolen} on which our analysis is
based, we arrived at closed equations by assuming equipartitioning
of the microscopic state probability within the observable
subshells. The resulting analytic description is in good agreement
with Langevin simulation for small and very large times, but not
sufficiently accurate for intermediate times. This is an artifact of
the equipartitioning ansatz and the truncation of the set of
observables to a relatively small (and computationally tractable)
set.

\section{Model Definitions}

The model consists of $N$ microscopic variables
$\bphi=(\phi_1,\ldots,\phi_N)$ representing the angular phase of
oscillators relative to a fixed frame of reference. At  equilibrium
the system is described by the Hamiltonian
\begin{equation}
H(\bphi)=-J\sum_{(i,j)\in\mathcal{G}_N}L_p\left(\cos(\phi_i-\phi_j-\omega_{ij})\right)
\label{eq:H}
\end{equation}
where $\mathcal{G}_N=(\mathcal{V}_N,\mathcal{E}_N)$ denotes a graph
instance of a set of vertices $\mathcal{V}_N=\{1,\ldots,N\}$ and
edges $\mathcal{E}_N=\{(i,j)|i,j\in\mathcal{V}_N\}$. The function
$L_p(x)$ denotes a $p$-th order Legendre polynomial. The thermal
variables take values from the interval $\phi_i\in[0,2\pi)$ for all
$i=1,\ldots,N$. We also introduced the `disorder' angle variables
$\omega_{ij}\in[0,2\pi)$ which represent locally preferred
orientations between sites $i$ and $j$. In a physical scenario these
are induced by the presence of impurities, cavities, or other
heterogeneities. We will assume that the distribution of these
variables is for all $i<j$ given by
\begin{equation}
K(\omega_{ij})=\frac12\,\delta_{\omega_{ij},\bar{\omega}}+\frac12\,\delta_{\omega_{ij},-\bar{\omega}}
\label{eq:K}
\end{equation}
for some $\bar{\omega}\in[0,2\pi)$. A more general treatment is
straightforward provided that the detailed balance condition  is
met. For each spin in the system we assign a local neighborhood,
i.e.\@ a set of sites to which the spin is connected. Let us for
simplicity abbreviate this set using $\partial i\equiv \{\ell|
(\ell,i)\in \mathcal{E}_N\}$, with $|\partial i|$ representing the
size  of the neighbourhood of $i$. Our graph is characterized by the
`degree' distribution
\begin{equation}
p(k)=\lim_{N\to\infty}\frac1N\sum_{i}\delta_{k,|\partial i|}
\label{eq:degree_dist}
\end{equation}
Moments of this distribution will be denoted $\bra k^m \ket =
\sum_{k\geq 0} p(k) k^m$. We draw graphs uniformly from the ensemble
of all graphs that have this given degree distribution, i.e.\@ we
use the configuration model \cite{Molloy-Reed}. In the thermodynamic
limit this leads to interactions on a sparse graph for which the
Bethe approximation is exact.

\section{Equilibrium analysis}

\subsection{The cavity formalism on a single instance}

To evaluate thermodynamic  properties and the phase diagram of this
system we have two main analytic tools at hand: the replica
\cite{MezardParisiVirasoro} and the cavity method \cite{mezard}.
Both of these have been used widely in a variety of settings, both
in infinite- and finite-connectivity systems. Although the
underlying philosophy of the two methods is different they deliver
identical results. In this paper we will choose to follow the cavity
method as developed for real-valued variables in
\cite{XYsmallworld}.

The starting point of this  method is to consider the marginal
probability of finding a spin at site $i$ in state $\phi_i$. This
follows from
\begin{equation}
P(\phi_i)=\frac{1}{Z_i}\int \d\phi_{\partial i}\, \e^{\beta
\sum_{\ell\in\partial i}L_p(\cos(\phi_i-\phi_\ell-\omega_{i\ell}))}
P^{(i)}(\phi_{\partial i}) \label{eq:true}
\end{equation}
where $Z_i$ is the appropriate normalisation constant.
$P^{(i)}(\phi_{\partial i})$ is the joint probability of finding the
neighbours of $i$ in state $\phi_{\partial
i}=(\phi_{j_1},\ldots,\phi_{j_{|\partial i|}})$ in the absence of
site $i$. To understand the origin of this expression we imagine an
iterative process that creates an exact tree structure in which each
spin is connected to $k$ neighbours apart from the site at the top
of the tree which has $k-1$ neighbours. At each step of this process
we add a new spin to the system by bringing together $k-1$
independent branches and connecting the top site of each of these to
the new site. Since, the $k-1$ branches influence each other only
through the new site that has been added, we can view the
probability $P^{(i)}(\phi_{\partial i})$ as the \emph{a priori}
probability of finding $\phi_{\partial i}$ before the addition of
the new spin, whereas the Boltzmann factor accounts for the
energetic cost of the merging. Since before the addition of the new
site into the system, the $k-1$ branches were independent we can
factorize $P^{(i)}(\phi_{\partial i})$:
\begin{equation}
P^{(i)}(\phi_{\partial i})=\prod_{\ell\in\partial i}P^{(i)}(\phi_\ell)
\label{eq:bethe}
\end{equation}
This equation is the Bethe approximation and is exact,  by
construction, on trees. For random graph structures where loops,
although predominantly long (e.g. scaling as $\log N$ on
Erd\"os-Reyni random graphs), do exist, (\ref{eq:bethe}) is
\emph{approximately} correct.

Equation (\ref{eq:true}) is not closed as it relates the \emph{true}
probability distribution with the \emph{cavity} probability
distribution. To close it we simply remove a neighbour of $i$ which
leads to
\begin{equation}
P^{(j)}(\phi_i)=\frac{1}{Z_i^{(j)}}\int \d\phi_{\partial i\setminus j}\,
\e^{\beta \sum_{\ell\in\partial i\setminus j}
L_p(\cos(\phi_i-\phi_\ell-\omega_{i\ell}))} \prod_{\ell \in\partial
i\setminus j} P^{(i)}(\phi_{\ell})
\label{eq:cavity}
\end{equation}
The two types of distributions that appear above can be dealt with
either (i) by an explicit parametrization, namely setting
$P^{(j)}(\phi_i)\to P(\phi_i|\{\mu_i^{(j)}\})$ (and similarly for
$P(\phi_i)$) where $\{\mu_i^{(j)}\}$ play the role of cavity fields
\cite{XYsmallworld,CoolenXY} or (ii) by using a simple histogram
for each of the distibutions. Note that since spins are continuous
variables, we are in principle required to introduce an infinite
number of cavity fields. For all practical purposes however, one
truncates the number of fields, while in some cases appropriate
choices for the parameterization can be made which ensure that the
impact of the truncation is not significant \cite{CoolenXY}. The
histogram method  of course requires advanced computational power
since one needs to allocate memory space for a relatively large
number of bins, but this approach has the advantage of working
directly with the distributions.

Once the stationary values of (\ref{eq:cavity}) are known for all
$i=1,\ldots,N$ and $j\in\partial i$ we can evaluate the true
probability function (\ref{eq:true}) and subsequently observables in
the system, e.g. the `magnetizations'
\begin{equation}
\left(\begin{array}{c} m_c \\ m_s \end{array}\right)
=\lim_{N\to\infty}\frac{1}{N}\sum_i\int \d\phi_i\ P(\phi_i)
\left(\begin{array}{c}\cos(\phi_i) \\ \sin(\phi_i)\end{array}\right)
\end{equation}
and similarly for other order parameters.

\subsection{The cavity formalism in the graph ensemble}

To perform a bifurcation analysis of this system it is convenient to
work with a graph ensemble, i.e.\@ the set of all instance graphs
$G_N=\{\mathcal{G}_N\}$ with the given degree distribution
(\ref{eq:degree_dist}).

We now consider the population of the cavity probability
distributions in the graph for all sites. This defines a functional
density in the following way
\begin{equation}
W[\{P\}]=\lim_{N\to\infty}\frac1N\sum_{i}\frac{1}{|\partial
i|}\sum_{j\in\partial i}\delta_{(F)}\left[P(\phi)-P_i^{(j)}(\phi)\right]
\label{eq:W_def}
\end{equation}
where  by $\delta_{(F)}$ we mean a functional delta distribution in
the sense that  $w[f]=\int dg\,w[g]\,\delta_{(F)}[g(x)-f(x)]$.
Similarly, for the density of true probability distributions
\begin{equation}
W_{\rm
true}[\{P\}]=\lim_{N\to\infty}\frac1N\sum_i\delta_{(F)}\left[P(\phi)-P_i(\phi)\right]
\label{eq:Wtrue_def}
\end{equation}
We can now convert (\ref{eq:true}) and (\ref{eq:cavity}) into
self-consistent equations for (\ref{eq:W_def}) and
(\ref{eq:Wtrue_def}) for the ensemble of graphs with degree
distribution $p(k)$
\begin{eqnarray}
W[\{P\}]&=&\sum_{k\geq 0}\frac{p(k)k}{\bra k\ket}\, \int
\prod_{\ell=1}^{k-1}
[\{\d P_\ell\}  W[\{P_\ell\}]\,\d\omega_\ell\,K(\omega_\ell)]\nonumber
\\
& &
\times\
\delta_{(F)}\left[P(\psi)-\frac{1}{\mathcal{Z}}\,\prod_{\ell=1}^{k-1}
\int \rmd \psi'\ P_\ell(\psi')\ \e^{\beta
JL_p\left(\cos(\psi-\psi'-\omega_{\ell})\right)}\right]
\label{eq:self_W}
\end{eqnarray}
where $\mathcal{Z}$ is the appropriate normalization constant. The
prefactor $p(k)k/\bra k\ket$ expresses the fact that the probability
that any given site is chosen to become a cavity site is
proportional to the number of bonds it has. For the density of true
probability distributions we have
\begin{eqnarray}
W_{\rm true}[\{P\}]&=&\sum_{k\geq 0}p(k)\, \int \prod_{\ell=1}^{k}
[\{\d P_\ell\}  W[\{P_\ell\}]\,\d\omega_\ell\,K(\omega_\ell)]\nonumber
\\
& &
\times\ \delta_{(F)}\left[P(\psi)-\frac{1}{\mathcal{Z}_{\rm
true}}\,\prod_{\ell=1}^{k}
\int \rmd \psi'\ P_\ell(\psi')\ \e^{\beta
JL_p\left(\cos(\psi-\psi'-\omega_{\ell})\right)}\right]
\label{eq:self_W}
\end{eqnarray}
Notice that so far we did not need to make a choice for the order of
the Legendre polynomial. One special case of the above equations is
in the absence of angular disorder and for a fixed degree
distribution $p_k=\delta_{k,c}$ and $K(\omega) = \delta(\omega)$. In
this case we have a $c$-regular graph in which virtually every spin
is living in an identical environment and one solution of
(\ref{eq:self_W}) is $W[\{P\}]=\delta[P(\phi)-P_\star(\phi)]$ for
some $P_\star(\phi)$. In the case where the order of the Legendre
polynomial is $p=1$, so that $L_1(x)=x$ and the degree distribution
is Poisson, our equations reduce identically to those of
\cite{CoolenXY} derived via the replica formalism.

Equation (\ref{eq:self_W}) can be solved in a spirit similar to the
`population dynamics' method \cite{mezard,CoolenXY}. As discussed in
the previous section, since $W[\{P\}]$ is  a measure over
distributions we can encode each of these distributions using a
simple histogram. Once a stationary solution for the $W[\{P\}]$ has
been obtained one can proceed to evaluate observables. For example,
observables describing magnetization and spin-glass order
respectively follow from
\begin{eqnarray}
m_c^{(k)}&=&\int \{\d F\}W_{\rm true}[\{P\}] \int \d \phi P(\phi)
\cos(k\phi) \label{eq:mc}\\
q_c^{(k)}&=&\int \{\d F\}W_{\rm true}[\{P\}] \left[\int \d \phi P(\phi)
\cos(k\phi)\right]^2 \label{eq:qc}
\end{eqnarray}
The order parameters
(\ref{eq:mc},\ref{eq:qc}) have the physical meaning
\begin{eqnarray}
m_{c}^{(k)}=\lim_{N\to\infty}\frac1N\sum_i\overline{\Bra
\cos(k\phi_i)\Ket_{eq}}\\
q_{c}^{(k)}=\lim_{N\to\infty}\frac1N\sum_i\overline{\Bra
\cos(k\phi_i)\Ket_{eq}^2}
\end{eqnarray}
where $\bra \cdots\ket_{eq}$ denotes thermal averages (similarly for
$m_{s}^{(k)}$ and $q_s^{(k)}$ as averages over $\sin(\phi)$).
Generally, the relevant set of observables depends (among other
factors) on the particular value of $p$. This is accounted by the
dependence of the above on the variable $k=1,\ldots,p$. One can
combine the above observables into a single pair e.g. via
\begin{equation}
m^{(k)}=\sqrt{(m_c^{(k)})^2+(m_s^{(k)})^2}\hspace{5mm}
q^{(k)}=\frac12(q_{c}^{(k)}+q_{s}^{(k)}) \label{eq:m}
\end{equation}
For  even values of $p$ there is no net magnetization in the system,
i.e. $m^{(1)}=0$, since the Hamiltonian is then rotationally
invariant via a uniform rotation of all spins through $\pi$.

\subsection{Bifurcation analysis}

To perform a bifurcation analysis, we notice that
$W[\{P\}]=\delta_{(F)}\left[P(\phi)-\frac{1}{2\pi}\right]$ satisfies
(\ref{eq:self_W}) for all temperatures. Thus, we can associate this
state with the high-temperature paramagnetic one (P). If one assumes
that bifurcations away from this solution occur in a continuous
fashion, then we can apply the so-called Guzai expansion
\cite{CoolenXY}, i.e.\@ consider the perturbation $P(\phi)\to
\frac{1}{2\pi}+\Delta(\phi)$, with $\int \rmd\phi \Delta(\phi)=0$
due to normalization. A linear stability analysis then produces the
following two conditions where, respectively, the first and second
moments of $W[\{P\}]$ bifurcate:
\begin{eqnarray}
1 &=&\frac{\bra k^2\ket-\bra k\ket}{\bra k\ket}
\underset{\ell>0}{\rm max}\bra\cos(\ell\omega)\ket_\omega\,
\frac{\mathcal{F}_p^{(\ell)}}{\mathcal{F}_p^{(0)}} \label{eq:PF}
\\
1 &=&\frac{\bra k^2\ket-\bra k\ket}{\bra
k\ket}\underset{\ell'>0}{\rm
max}\frac{(\mathcal{F}_p^{(\ell')})^2}{(\mathcal{F}_p^{(0)})^2}
\label{eq:PSG}
\end{eqnarray}
where
\begin{equation}
\mathcal{F}_p^{(\ell)}\equiv \int_0^{2\pi}\d \phi\,\cos(\ell\phi)\
\e^{\beta JL_p\left(\cos\phi\right)}
\end{equation}
(details on similar calculations can be found in  \cite{CoolenXY}).
These two conditions provide the critical temperatures where a
ferromagnetic (F) and a spin-glass phase (SG) appear, respectively.
The range of integers over which we maximize
(\ref{eq:PF},\ref{eq:PSG}) expresses physically the fact that at the
moment of bifurcation towards an ordered phase the distribution of
spin orientations has $\ell$ maxima (in fact possible bifurcating
modes occur at the Fourier modes $\cos(\ell\phi)$). For $p=2$ we
find that the bifurcating mode is given by $\ell=2$  implying that
spins can be ordered in a parallel or in an anti-parallel fashion
(they are energetically equivalent). For larger values of $p$ we
find a transition in which the bifurcation mode changes from
$\ell=2$ to $\ell=p$.

\begin{figure}[t]
\setlength{\unitlength}{0.1cm}
\begin{picture}(150,45)
\put(0,0){\epsfysize=50\unitlength\epsfbox{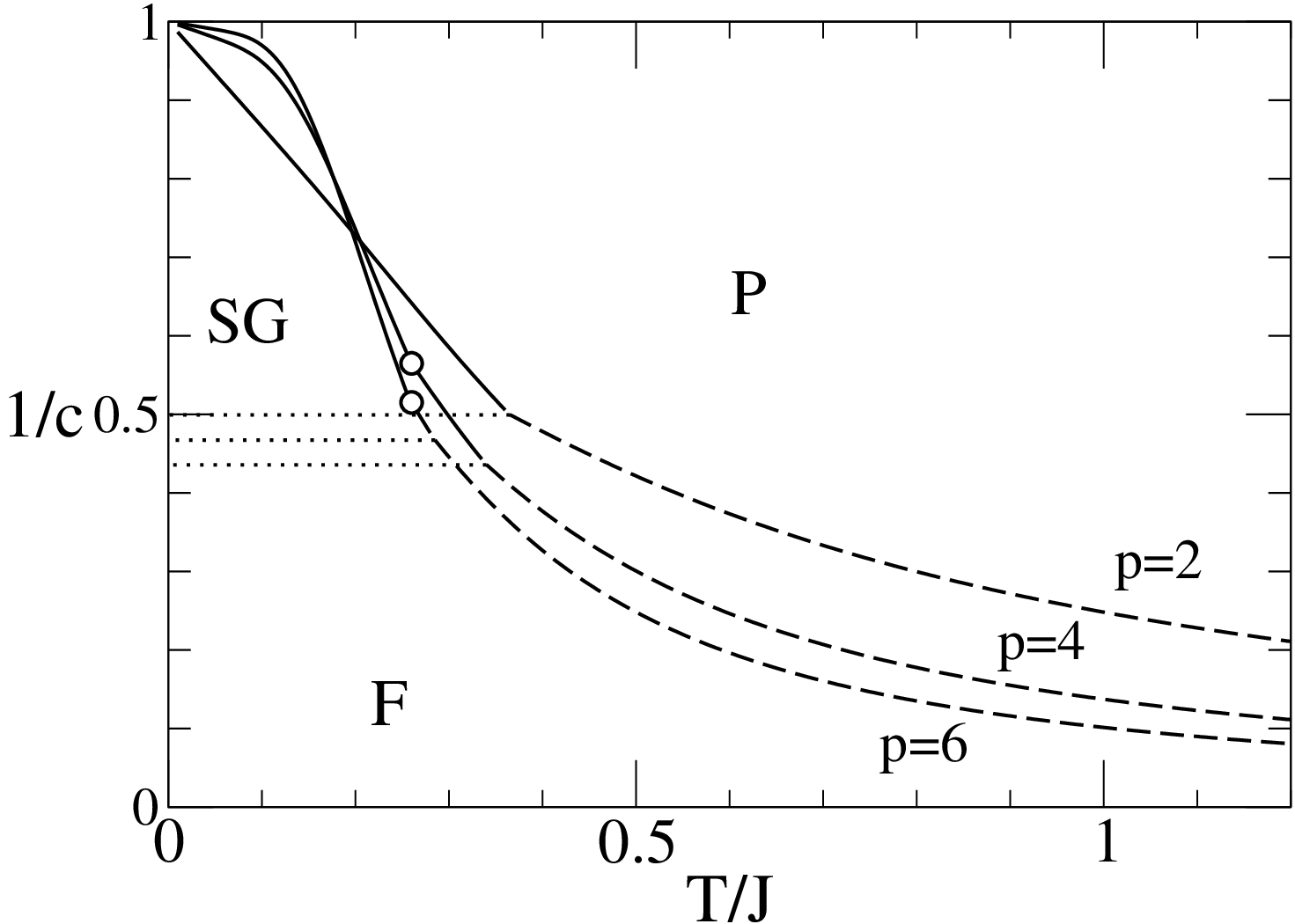}}
\put(75,0){\epsfysize=50\unitlength\epsfbox{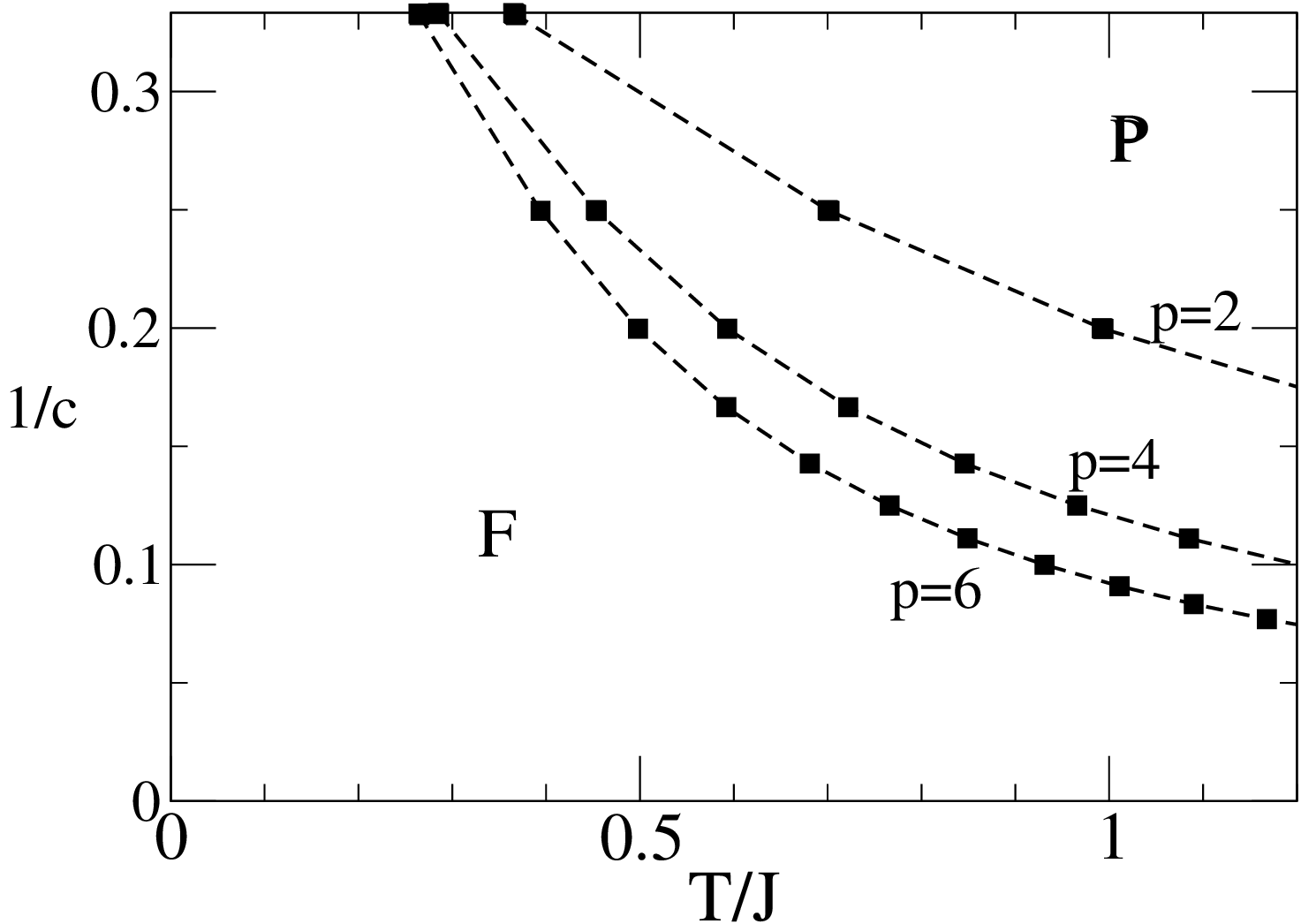}}
\end{picture}
\vspace{0mm} \caption{Phase diagrams of the model (\ref{eq:H}) for
$\bar{\omega}=\pi/8$ and for different values of $p$. Left figure:
Poisson graph with $p(k)=e^{-c}c^k/k!$.  Dashed, solid and dotted
lines represent the P$\to$F, P$\to$SG and F$\to$SG transition,
respectively. Open circles indicate a change in the bifurcating
mode. All transitions are second-order. Right figure: $c$-regular
graph; markers correspond to integer values of $c$ and lines are
guide to the eye. For $c = 3$ and $p = 4,6$ the transition is from
P$\to$SG.} \label{fig:poisson_omega0}
\end{figure}

\begin{figure}[t]
\setlength{\unitlength}{0.1cm}
\begin{picture}(160,55)
\put(0,0){\epsfysize=50\unitlength\epsfbox{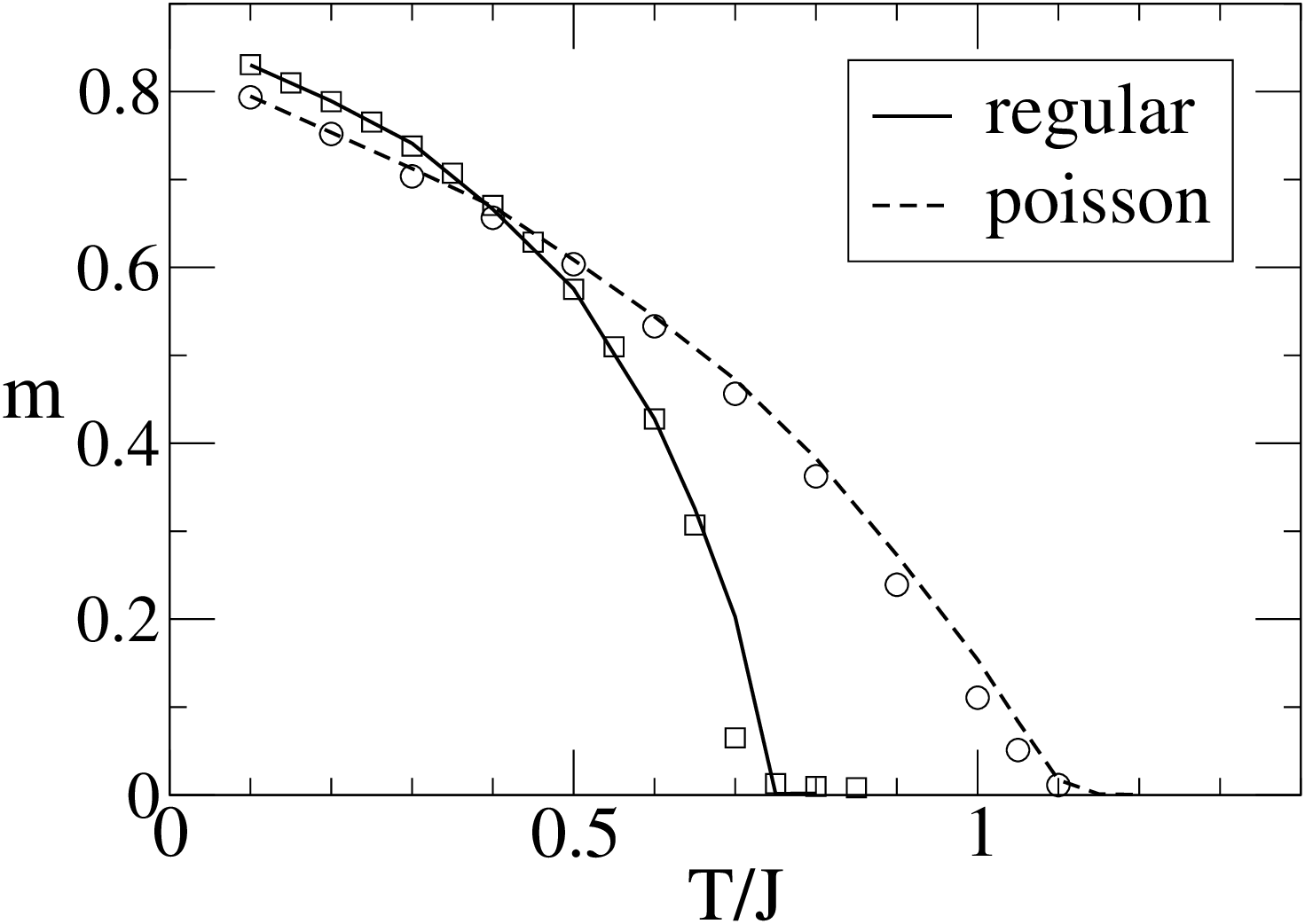}}
\put(75,0){\epsfysize=50\unitlength\epsfbox{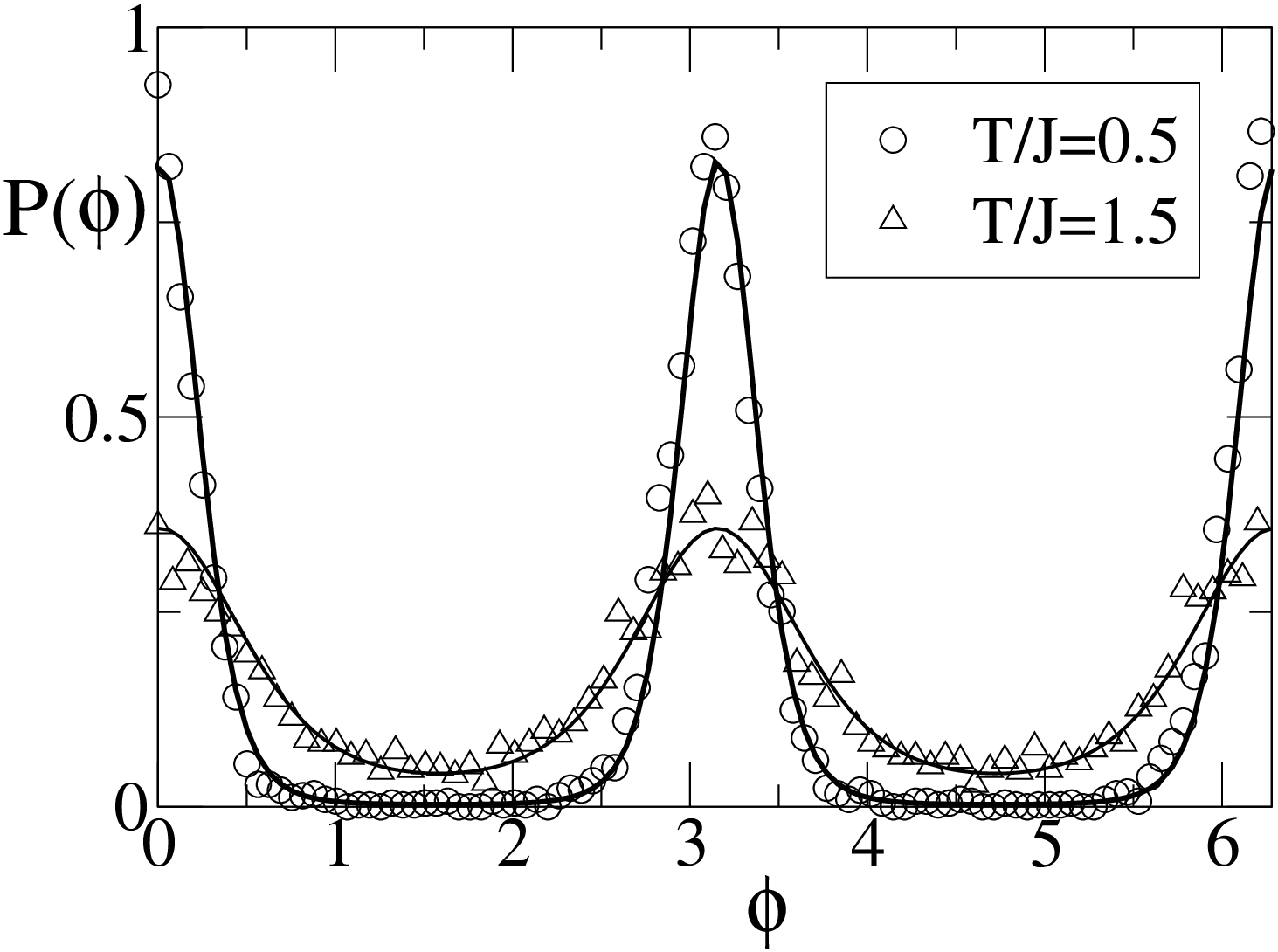}}
\end{picture}
\caption{Left: The order parameter $m^{(2)}$ (\ref{eq:m}) for the
$p=2$ model with $\bar{\omega}=\pi/8$ and for a Poisson- (dashed)
and a regular graph (solid) with $\bra k\ket=4$. The critical
temperatures agree well with the results of the bifurcation
analysis. Right: The distribution of rod orientations $P(\phi)$ for
$p=2$. We observe two peaks at $\phi=0,\pi$ reflecting the energetic
equivalence of parallel and anti-parallel alignment. Markers in both
pictures
correspond to simulation experiments.} \label{fig:m}
\end{figure}

Let us now describe the resulting phase diagrams in more detail. For
$\bar{\omega}=0$ one finds a phase diagram with a P$\to$F transition
only, while for $\bar{\omega}>0$ richer phase diagrams occur. In
figure \ref{fig:poisson_omega0} we plot phase diagrams for
$\bar{\omega}=\pi/8$ for different values of $p$. The left picture
corresponds to Poisson degree distributions $p(k)=e^{-c}c^k/k!$,
while in the right picture we show the phase diagram for a regular
graph. Larger values of $p$ produce sharper P$\to$F and P$\to$SG
transitions. The critical temperatures provided by simulation
experiments and the numerical solution of our equations (\ref{eq:m})
are in good agreement with the results of the bifurcation analysis
supporting the fact that all transitions are second-order (at least
for $p\leq 6$ where we have currently focused). For the F$\to$SG
transition we have assumed that it is given by the dotted line. This
follows from physical reasoning (absence of re-entrance phenomena
\cite{parisi}). The change in the bifurcation mode is given by the
open circles, namely for $T<T_{\rm circle}$ we find that $\ell=2$
solves the maximization problem in (\ref{eq:PF},\ref{eq:PSG}),
whereas for $T>T_{\rm circle}$ we have $\ell=p$. In the left picture
of figure \ref{fig:m} we show the magnetization order parameter
$m^{(2)}$ (\ref{eq:m}) for Poisson and regular random graphs of mean
connectivity 4. In the Poisson graph transitions are smoother due to
the variable number of connections per site. Markers correspond to
simulation experiments of $N=25,000$. In the right picture we show
the predicted distribution of rod orientations
$\mathcal{P}(\phi)=\int \{dP\}W_{\rm true}[\{P\}] P(\phi)$ against
simulation experiments with $\mathcal{P}(\phi)=\frac1N\sum_i\bra
\delta[\phi-\phi_i]\ket_{eq}$ for a $p=2$ model on a Poisson graph
with $\bra k\ket=5$, $\bar{\omega}=0$ and $T=0.5$ and $T=1.5$. We
observe that $P(\phi)$ has two peaks separated by $\pi$ which
reflects the fact that parallel and antiparallel orientations are
equivalent. Note that we have given an (arbitrary) overall rotation
to the spins so that the peaks of the distribution are aligned at 0
and $\pi$. Lower temperatures promote order and as a result
$P(\phi)$ is sharper for $T/J=0.5$ than for $T/J=1.5$.

\begin{figure}[t]
\setlength{\unitlength}{0.1cm}
\begin{picture}(0,55)
\put(10,-0){\epsfysize=45\unitlength\epsfbox{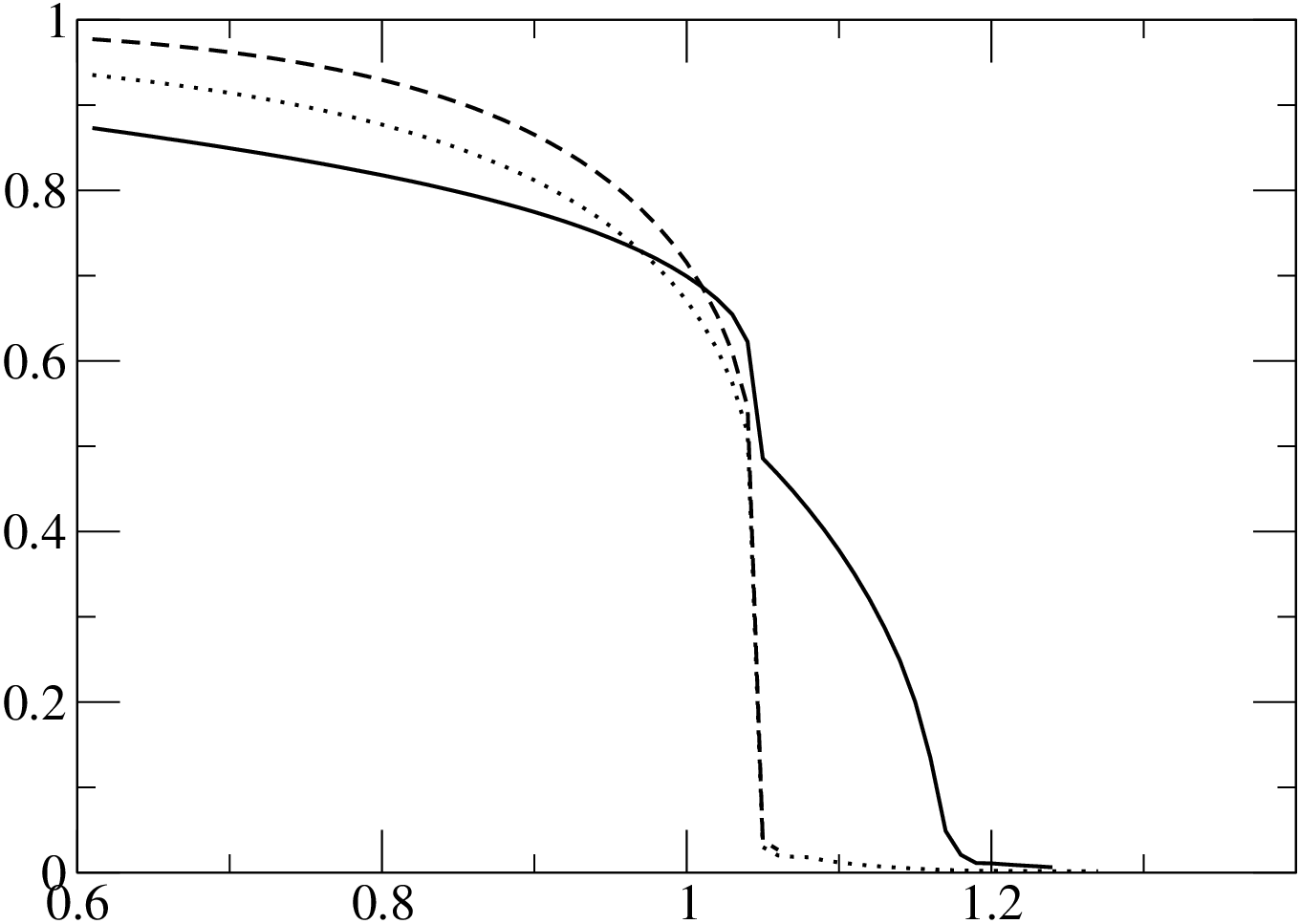}}
\put(4,28){\small $m$}
\put(45,-4){\small $T/J$}
\end{picture}
\vspace{5mm} \caption{The magnetisations $m^{(2)}$ (dashed), $m^{(4)}$ (dotted) and $m^{(6)}$
(solid), given by equation (\ref{eq:m}), for a system described by a Legendre order $p=6$ on a
regular lattice and with $\bar{\omega}=0$.  We see that there are two critical temperatures in this system.
 At the first one $T_{\rm crit}^{(1)}\approx 1.16$, which agrees with the prediction of (\ref{eq:PF}), only $m^{(6)}$ becomes non-zero.
 For temperature values $T<T_{\rm crit}^{(2)}\approx 1.04$ the magnetisations
$m^{(2)}$ and $m^{(4)}$ also become non-zero. This latter transition appears to be discontinuous.
}
\label{fig:p6}
\end{figure}

Finally, in figure \ref{fig:p6} we show the magnetizations
$m^{(2)}$, $m^{(4)}$ and $m^{(6)}$ for a system on a regular lattice
with $\bar{\omega}=0$. Here we have chosen $p=6$ to examine the
impact of choosing a high-order Legendre polynomial. We observe that
at the critical temperature $T_{\rm crit}^{(1)}\approx 1.16$ the
magnetisation $m^{(6)}$ becomes non-zero; while the other two
magnetisation order parameters remain zero. This bifurcation occurs
at the point predicted by (\ref{eq:PF}) which indicates that despite
the high $p$-value the transition is continuous. It is interesting
to note that the magnetisations $m^{(2)}$ and $m^{(4)}$ do not
bifurcate until a lower temperature $T_{\rm crit}^{(2)}\approx 1.04$
is reached. This transition appears to be discontinuous and, as it
is preceded by the transition at $T_{\rm crit}^{(1)}$, we have been
unable to derive it analytically (since in the ordered phase we are
required to find the distribution $P(\phi)$ numerically). For
temperature values between the two critical ones the ferromagnetic
phase describes a system with `local' order, i.e.\@ relative only to
the bifurcating mode (here, the sixth Fourier mode (\ref{eq:PF})).

\section{Dynamics}

We now turn to the relaxational dynamics of our system. The
microscopic dynamics of our model is dictated by the Langevin
equation
\begin{equation}
\frac{\d}{\d t}\phi_i(t)=-\frac{\partial H(\bphi)}{\partial
\phi_i}+\eta_i(t)
\label{eq:langevin}
\end{equation}
where $\eta_i(t)$ represents Gaussian white noise  $\bra
\eta_i(t)\ket=0$ and $\bra
\eta_i(t)\eta_j(t')\ket=2T\delta_{ij}\delta(t-t')$. Defining the
microscopic state probability $p_t(\bphi)=\bra
\delta(\bphi-\bphi(t))\ket$ (where $\bra \cdots\ket$ represents
average over the stochastic process) we can obtain from
(\ref{eq:langevin}) the corresponding Fokker-Planck  equation
\begin{equation}
\frac{\d}{\d t}p_t(\bphi)=\sum_{i=1}^N\frac{\partial}{\partial \phi_i}
\left[p_t(\bphi)\frac{\partial H(\bphi)}{\partial
\phi_i}\right]+T\sum_{i=1}^N\frac{\partial^2}{\partial
\phi_i^2}p_t(\bphi) \label{eq:fokkerplanck1}
\end{equation}
We  now consider a set of $\ell$ macroscopic observables which we
denote collectively as
$\bOmega(\bphi)=(\Omega_1(\bphi),\ldots,\Omega_\ell(\bphi))$ and we
define accordingly the macroscopic probability distribution for
these observables:
\begin{equation}
\mathcal{P}_t(\bOmega)=\int
\d\bphi\,p_t(\bphi)\,\delta[\bOmega-\bOmega(\bphi)]
\end{equation}
From (\ref{eq:fokkerplanck1}) one can derive, through integration
by-parts, a Fokker-Planck equation for $\mathcal{P}_t(\bOmega)$
\begin{eqnarray}
\frac{\d}{\d
t}\mathcal{P}_t(\bOmega)=-\sum_{k=1}^\ell\frac{\partial}{\partial
\Omega_k} \left[\mathcal{P}_t(\bOmega)\Bra \sum_i \frac{\partial
H}{\partial \phi_i}\frac{\partial \Omega_k}{\partial
\phi_i}-T\frac{\partial^2\Omega_k} {\partial
\phi_i^2}\Ket_{\bOmega,t}\right]\nonumber
\\
+T\sum_{h,k=1}^\ell\frac{\partial^2}{\partial \Omega_h\partial
\Omega_k}\left[\mathcal{P}_t(\bOmega)\Bra \sum_i \frac{\partial
\Omega_h}{\partial \phi_i}\frac{\partial \Omega_k}{\partial
\phi_i}\Ket_{\bOmega,t}\right] \label{eq:fokkerplanck2}
\end{eqnarray}
which is expressed in terms of the so-called subshell averages
\begin{equation}
\Bra [\cdots ]\Ket_{\bOmega,t}\equiv\frac{\int
\d\bphi\,p_t(\bphi)\,\delta[\bOmega-\bOmega(\bphi)]\,[\cdots]} {\int
\d\bphi\,p_t(\bphi)\,\delta[\bOmega-\bOmega(\bphi)]}
\end{equation}
Using (\ref{eq:fokkerplanck2}) we obtain the evolution equations for
the macroscopic observables. Let us denote $\Omega_k(t)=\int
\d\bOmega\,\mathcal{P}_t(\bOmega)\, \Omega_k$. Then integration of
(\ref{eq:fokkerplanck2}) gives
\begin{equation}
\frac{\d}{\d t}\Omega_k(t)= \sum_i\Bra \frac{\partial H}{\partial
\phi_i}\frac{\partial \Omega_k}{\partial
\phi_i}-T\frac{\partial^2\Omega_k} {\partial
\phi_i^2}\Ket_{\bOmega_t,t} \label{eq:evolution}
\end{equation}
(assuming  that the relevant boundary terms vanish) with the
notation $\bra [\cdots]\ket_{\bOmega_t,t}=\int
d\bOmega\,p_t(\bOmega)\,\Bra[\cdots]\Ket_{\bOmega,t}$. Note that
(\ref{eq:evolution}) while exact (for reasonable choices of
$\bOmega$) is not closed as it still depends on the microscopic
probability $p_t(\bphi)$. To eliminate this dependence we now make
the equipartitioning assumption which underlies dynamical replica
theory \cite{Coolen}: we assume that $p_t(\bphi)$ depends on $\bphi$
only through $\bOmega(\bphi)$ so that $p_t(\bphi)$ can be removed
completely from the subshell average (i.e. all microstates which
give the same value of the macroscopic observables are equally
likely).

Let us now  be more specific and choose as our observables the
magnetization and energy
\begin{eqnarray}
m_c^{(k)}&=&\frac1N\sum_{i}\cos(k\phi_i)
\hspace{10mm}
e=-\frac{J}{N}\sum_{(i,j)\in\mathcal{G}}L_p\left(\cos(\phi_i-\phi_j-\omega_{ij})\right)
\\
m_s^{(k)}&=&\frac1N\sum_{i}\sin(k\phi_i)
\end{eqnarray}
For notational convenience we will from now suppress the time
dependence in the above i.e.\@ we will write $m^{(k)}_{c,t}\to
m^{(k)}_c$ and similarly for the other observables. Calculating  the
various derivatives in (\ref{eq:evolution}) and inserting into the
equations the joint spin-field distribution
\begin{eqnarray}
\lefteqn{D_{m^{(k)}_{c},m^{(k)}_{s},e}(h,\phi)=}\nonumber
\\
& &=\frac1N\Bra\sum_i\delta\left[h-\sum_{\ell\in\partial
i}\sin(\phi-\phi_\ell-\omega_{i\ell})
\frac{\partial L_p(x)}{\partial x}\right]\delta(\phi-\phi_i)
\Ket_{m^{(k)}_{c},m^{(k)}_{s},e,t}
\end{eqnarray}
results in the trio of ordinary differential equations
\begin{eqnarray}
\frac{\d}{\d t}m^{(k)}_{c}
& = &
Tk^2 m^{(k)}_{c}-Jk\int \d h\d
\phi\,D_{m^{(k)}_{c},m^{(k)}_{s},e}(h,\phi)\,h\sin(k\phi)
\\
\frac{\d}{\d t}m^{(k)}_{s}
& = &
Tk^2m^{(k)}_{s}-Jk\int \d h\d
\phi\,D_{m^{(k)}_{c},m^{(k)}_{s},e}(h,\phi)\,h\cos(k\phi)
\\
\frac{\d}{\d t}e
& = &
-2Te+J\int \d h\d \phi\,D_{m^{(k)}_{c},m^{(k)}_{s},e}(h,\phi)\,h
\end{eqnarray}
The distribution $D_{m^{(k)}_{c},m^{(k)}_{s},e}(h,\phi)$ represents
the distribution of spins and local fields in the
\emph{microcanonical} ensemble for a given value of the
magnetization and energy. Once we know this  at time step $t$ we can
predict the values of the observables at time step $t+\Delta t$. To
calculate this joint distribution our strategy will be to associate
it to a quantity with the same physical meaning but in another
physical setting \cite{SemerjianWeigt}: we consider a
\emph{canonical} ensemble where the temperature and external fields
will force the system to have the values $m^{(k)}_{c/s,t}$ and
$e_t$. In this new system the Boltzmann distribution is given by
\begin{equation}
P(\bphi)=\frac{1}{\mathcal{Z}(\gamma,\{h_c^{(k)},h_s^{(k)}\})}
\e^{-\gamma H(\bsigma)-\gamma \sum_{k\geq
1}[h_c^{(k)}\sum_{i}\cos(k\phi_i)+h_s^{(k)}\sum_i\sin(k\phi_i)]}
\end{equation}
while the joint spin-field distribution can be defined as
\begin{equation}
\mathcal{D}(h_i,\phi_i)=\int \d \phi_{\partial i}\,Q(\phi_i,\phi_{\partial
i})\,
\delta\left[h_i-\sum_{\ell\in\partial
i}\sin(\phi_i-\phi_\ell-\omega_{i\ell})\frac{\partial L_p(x)}{\partial
x}\right]
\label{eq:joint_canonical}
\end{equation}
where $Q(\phi_i,\phi_{\partial i})$ represents the joint
distribution of finding spin $i$ and its neighbours in the given
state. Using the reasoning that follows equation (\ref{eq:true}) and
the Bethe approximation (\ref{eq:bethe}) we can write for tree-like
structures
\begin{equation}
Q(\phi_i,\phi_{\partial i})\sim\e^{\gamma \sum_{k\geq
1}[h_c^{(k)}\cos(k\phi_i)+h_s^{(k)}\sin(k\phi_i)]+
\gamma J\sum_{\ell\in\partial
i}L_p(\cos(\phi_i-\phi_j-\omega_{i\ell}))}\,\prod_{\ell\in\partial
i}P^{(i)}(\phi_{\ell})
\end{equation}
with $P^{(j)}(\phi_i)$, as before, representing the cavity
distribution of finding spin $i$ in $\phi_i$ in the absence of site
$j$ and in the symbol $\sim$ we have absorbed the normalization.
From here, integration with respect to the spin variables of the
neighborhood of $i$ gives the marginal distribution
\begin{equation}
P(\phi_i)\sim\e^{\gamma \sum_{k\geq
1}[h_c^{(k)}\cos(k\phi_i)+h_s^{(k)}\sin(k\phi_i)]}\int \d \phi_{\partial i}
\e^{\gamma J\sum_{\ell\in\partial
i}L_p(\cos(\phi_i-\phi_j-\omega_{i\ell}))}\,\prod_{\ell\in\partial
i}P^{(i)}(\phi_{\ell})
\end{equation}
and removing a neighbour of $i$ closes this to
\begin{eqnarray}
P^{(j)}(\phi_i)&\sim&
\e^{\gamma \sum_{k\geq 1}[h_c^{(k)}\cos(k\phi_i)+\gamma
h_s^{(k)}\sin(k\phi_i)]}\nonumber
\\
& &
\times\ \int \d \phi_{\partial i\setminus j}\e^{\gamma
J\sum_{\ell\in\partial i\setminus
j}L_p(\cos(\phi_i-\phi_j-\omega_{i\ell}))}\,\prod_{\ell\in\partial
i\setminus j}P^{(i)}(\phi_{\ell})
\end{eqnarray}
Observables in this canonical ensemble are given by
\begin{eqnarray}
\mu_c^{(k)}&=&\lim_{N\to\infty}\frac1N\sum_i\int \d \phi \ P(\phi_i)
\cos(k\phi_i)
\label{eq:mc_canonical}
\\
\mu_s^{(k)}&=&\lim_{N\to\infty}\frac1N\sum_i\int \d \phi \ P(\phi_i)
\sin(k\phi_i)
\label{eq:ms_canonical}
\\
\epsilon &=&-\lim_{N\to\infty}\frac{J}{N}\sum_{(i,j)\in\mathcal{G}_N}\Bra
L_p(\cos(\phi_i-\phi_j-\omega_{ij}))\Ket_{\star}
\label{eq:e_canonical}
\end{eqnarray}
with the short-hand notation
\begin{equation}
\Bra \cdots
\Ket_{\star}=\frac{\int\d\phi_i\d\phi_j\,P^{(i)}(\phi_j)P^{(j)}(\phi_i)\,
\e^{\gamma J L_p(\cos(\phi-\phi_j-\omega_{ij}))}[\cdots]}
{\int \d\phi_i\d\phi_j\,P^{(i)}(\phi_j)P^{(j)}(\phi_i)\,\e^{\gamma J
L_p(\cos(\phi_i-\phi_j-\omega_{ij}))}}
\end{equation}
The above equations can be easily re-written  in a graph-ensemble
form for a given degree distribution.

We see that this is now equivalent to an equivalent static
calculation (using a more involved Hamiltonian): at each time step
$t$ and given the values of observables $m^{(k)}_{c/s,t}$ and $e_t$,
one has to find the values of the temperature $\gamma$ and external
fields $h_c^{(k)},h_s^{(k)}$ that via the definitions
(\ref{eq:mc_canonical}), (\ref{eq:ms_canonical}) and
(\ref{eq:e_canonical}) ensure that the equalities
$\mu^{(k)}_{c}=m^{(k)}_{c,t}$, $\mu^{(k)}_s=m^{(k)}_{s,t}$,
$\epsilon=e_t$ are satisfied. Once the correct values of the
parameters have been found one can evaluate the joint spin-field
distribution (\ref{eq:joint_canonical}) and this corresponds to
$D_{m^{(k)}_{c},m^{(k)}_{s},e}(h,\phi)$.

\begin{figure}[h]
\setlength{\unitlength}{0.1cm}
\begin{picture}(150,50)
\put(0,0){\epsfysize=50\unitlength\epsfbox{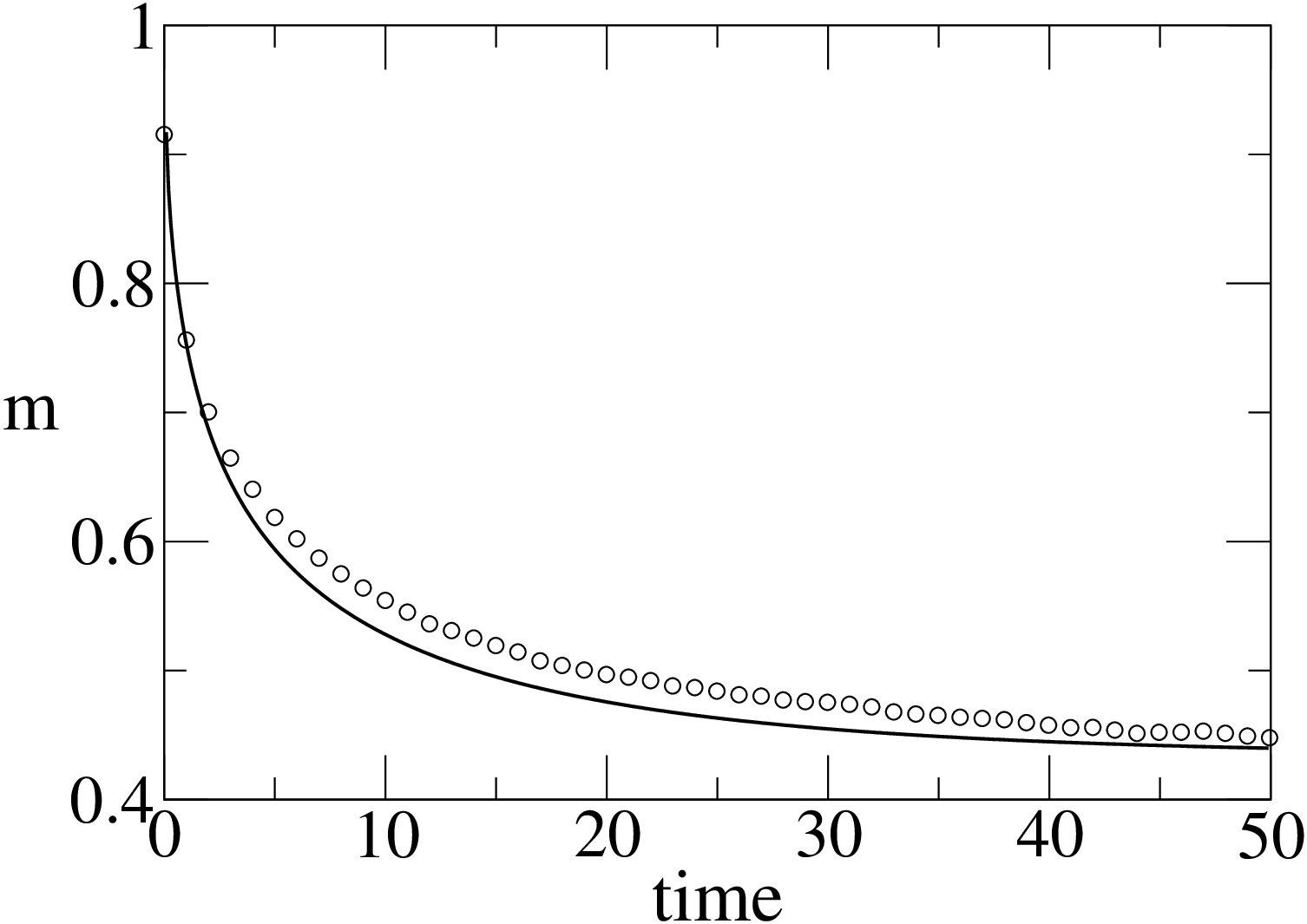}}
\end{picture}
\caption{The evolution of the magnetization
$m^{(2)}=[m_c^{(2)}+m_s^{(2)}]^{\half}$ for a 3-regular lattice and
$p=2$ from a highly ordered initial state in a heat-bath of
temperature $T=0.6$ and $\bar{\omega}=0$. The solid line represents
the theoretic prediction and markers Langevin simulations of $N =
40\, 000$ spins. } \label{fig:DRT}
\end{figure}

While the conceptual formulation of this methodology is relatively
straightforward, the numerical implementation is a bit more
challenging. For each time step, one has to solve the inverse
problem of finding the correct values of the temperature $\gamma$
and external fields $h_c^{(k)},h_s^{(k)}$ which produce a given
magnetization and energy. Thus the search space is already
$(2k+1)$-dimensional and  computational costs are not negligible.
The algorithm also requires the convergence of population dynamics
and we are thus required to work in regions of phase space where
only trivial ergodicity breaking occurs.  For these reasons we have
implemented our algorithm in the simple scenario of $p=2$ (in which
case $m^{(1)}=0$) which allows us to  restrict the present set of
observables to $(m_c^{(2)},m_s^{(2)},e)$. We have considered a
3-regular graph. The results are shown in figure \ref{fig:DRT}. We
see that the analytic solution compares perfectly with simulations
for small and very large times, but for intermediate times the two
clearly deviate. This is due to the assumptions we have made, namely
equipartitioning in the space of the three observables. We expect
that by choosing a larger set of macroscopic observables (e.g.\@
incorporating correlation functions) this approximation will
improve. There is also an error introduced due to the presence of
loops in the graph; for smaller graphs we would expect a larger
deviation due to the incorrectness of the Bethe ansatz, while in the
truly infinite system these will not occur.

\vspace{10mm}

\section{Discussion}

In this paper we have studied the Lebwohl-Lasher model, which has a
characteristic energy landscape of sharp and narrow wells. This
model has been studied extensively in liquid crystal models and our
approach here can be seen as the Bethe approximation to the finite
dimensional problem. We have considered random graph lattices. At
equilibrium our treatment follows the general methodology developed
in \cite{XYsmallworld} based on the cavity method within the ergodic
(replica-symmetric) assumption. We have given bifurcation conditions
describing the phase diagram. Numerical evaluation of the order
parameters shows the transition from ordered to paramagnetic phase.
This transition, as in Ising spin systems, is sharper for regular
than for random lattices. Simulation experiments are in excellent
agreement with the analytic results.

We have also extended the formalism developed in \cite{GuzaiDRT} for
the dynamics of spin systems on finitely-connected systems into
situations where spins are continuous variables. Our starting point
has been the Langevin equation for the microscopic dynamics from
which a set of differential equations follows for our chosen set of
observables. The key to solving this set of evolution equations is
the approximations of dynamical replica theory \cite{Coolen}. We
have contrasted the results of this methodology with simulation
experiments. This shows that despite the appealingly clean analytic
form, there are weaknesses of the formalism that need to be dealt
with in the future. In particular, the assumption of
equipartitioning of the microscopic state probability within the
subshells is too strong for small sets of observables while for
larger ones the computational costs start becoming prohibitive.

\section*{Acknowledgements}
We are indebted to Isaac P\'erez Castillo and Tatsuya Uezu with whom
we developed the dynamics of continuous spin systems on graphs. NS
wishes to thank A C D van Enter for a very motivating introduction
to liquid crystal models. This paper has been prepared by the
authors as a commentary on the topic as at November 2006 and is not
a definitive analysis of the subject matter covered.  It does not
constitute advice or guidance in this area and should not be relied
upon as such.  The paper has not been prepared by JH in his capacity
as an employee of Hymans Robertson LLP and the views expressed do
not represent those of Hymans Robertson LLP. Neither the authors nor
Hymans Robertson LLP accept liability for any errors or omissions.

\end{document}